\documentclass[a4paper]{article}

\usepackage{INTERSPEECH2021}
\usepackage{multirow}
\usepackage{amssymb}
\usepackage{adjustbox}
\usepackage{comment}

\title{Investigation of Speaker-adaptation methods in Transformer based ASR}
\name{Vishwas M. Shetty, Metilda Sagaya Mary N J, S. Umesh}

\address{
  Department of Electrical Engineering, Indian Institute of Technology Madras, Chennai, India}
\email{ee17s045@smail.iitm.ac.in, ee18d013@smail.iitm.ac.in, umeshs@ee.iitm.ac.in}

\begin{document}

\maketitle
\begin{abstract}
End-to-end models are fast replacing the conventional hybrid models in automatic speech recognition. Transformer, a sequence-to-sequence model, based on self-attention popularly used in machine translation tasks, has given promising results when used for automatic speech recognition. This paper explores different ways of incorporating speaker information at the encoder input while training a transformer-based model to improve its speech recognition performance. We present speaker information in the form of speaker embeddings for each of the speakers. We experiment using two types of speaker embeddings: \textbf{\textit{x-vectors}} and novel \textbf{\textit{s-vectors}} proposed in our previous work. We report results on two data-sets a) NPTEL lecture database and b) Librispeech 500-hour split. NPTEL is an open-source e-learning portal providing lectures from top Indian universities. We obtain improvements in the word error rate over the baseline through our approach of integrating speaker embeddings into the model.

\end{abstract}

\section{Introduction}
The recognition performance of an Automatic Speech Recognition (ASR) system is affected by speaker variations. Speaker adaptation in conventional DNN-HMM based systems was explored in \cite{DNN_adapt_seki,DNN_adapt_bell,ivec_Saon,ivec_gupta,ivec_andrew,cortana_speaker_adapt}. I-vectors appended to input features have been shown to improve the model performance. The use of i-vectors for speaker adaptation was explored in \cite{ivec_Saon,ivec_gupta,ivec_andrew}. Along with i-vectors, the use of x-vectors, deep CNN embeddings for speaker adaptive training of DNN was explored in \cite{DNN_adapt_bell}. In \cite{cortana_speaker_adapt}, d-vectors were used for speaker adaptation. 

Speaker adaptation in an End-to-End (E2E) framework was explored in \cite{ctc_adapt,hansen_e2e_adapt,Shinji_speaker_Adapt_e2e,moritz_e2e}. In \cite{moritz_e2e}, speaker i-vectors extracted from the training data were stored in a memory block and accessed through a learned attention mechanism during testing. The resulting vector was called the memory vector and was appended to the acoustic features while training the E2E model. This was an unsupervised speaker adaptation approach because it did not require i-vector computation during testing. In \cite{ctc_adapt}, two approaches for adaptation, Kullback-Leibler divergence (KLD) adaptation and Multi-task learning (MTL) adaptation of an E2E Connectionist Temporal Classification (CTC) based model was proposed. Speaker adaptation in a multi-channel E2E framework was proposed in \cite{Shinji_speaker_Adapt_e2e}.

In \cite{synthesis_Adapt_microsoft}, a method for speaker adaptation by using a personalized speech synthesizer and a neural language generator was proposed. The primary objective in \cite{synthesis_Adapt_microsoft} was to address data sparsity in the case of rapid speaker adaptation. Their approach was similar to back-translation in a machine translation problem. The speech synthesizer was adapted to the target speaker by using about a minute of speech data, which was then used to synthesize relevant text generated by the neural language generator. The synthesized speech and the original speech data for adaptation were then used for the acoustic model adaptation. 

In the transformer-based E2E framework, \cite{bo_xu_transformer} proposed a speaker-aware speech-transformer. Here speaker embeddings were obtained by attention over i-vectors. At each time step, a weighted combination of i-vectors was calculated to generate a speaker embedding.

In this paper, we propose to study the effect of providing speaker information on ASR performance for systems built on a transformer framework. We explore different ways of providing speaker information while training the model as well as during testing. Speaker information is provided in the form of embeddings for each of the speakers. We have experimented with two different types of embeddings:  \textbf{\textit{x-vectors}} and our newly introduced \textbf{\textit{s-vectors}}. We have also explored different ways of incorporating them: \textit{add} and \textit{concatenate}. Results are reported both in seen and unseen speaker scenarios.

\section{Different Speaker Embeddings}

The \textit{Baseline} model in our experiment was the model trained without speaker information provided in any form. This speaker-independent model was trained with 83-dimensional features, i.e., 80-dimensional filter bank features and three pitch-related features. These were obtained by using 25ms window frames,  with a frame shift of 10ms. These were then passed through two conv2d layers with stride two each. Hence the input of \textit{T} frames to the conv2d  layers was reduced to \textit{T/4} at the output of conv2d. 
For speaker adaptation, we provide  speaker embeddings both during training and testing, i.e., speaker-adaptive training. In the following sections, we describe the different ways in which we provide the speaker embeddings.
\vspace{-1.0em}
\subsection{x-vectors as embeddings}
A 512 dimension x-vector embedding is obtained for each speaker using a pre-trained model built-in Kaldi. The pre-trained model\footnote{http://kaldi-asr.org/models/8/0008\_sit\_v2\_1a.tar.gz} was built using Voxceleb data. The \textbf{\textit{x-vectors}} are extracted from Time Delay Neural Network (TDNN) based architecture. We have used \textbf{\textit{x-vectors}} extracted per speaker while training our models; for example, if we have “N” speakers, we extract “N” \textbf{\textit{x-vectors}}.
\vspace{-1.0em}
\subsection{s-vectors as embeddings}
S-vector \cite{metilda_speaker_verification} proposed by us has outperformed the \textbf{\textit{x-vector}} on both Voxceleb-1 and Voxceleb-2  Speaker Verification task by a significant margin. \textbf{\textit{S-vectors}} are extracted from transformer encoder based architecture. The model is first trained for speaker recognition, and then the corresponding vectors are obtained by tapping the models at the feed-forward layer. Self-attention, is the backbone of \textbf{\textit{s-vectors}} and can capture better speaker characteristics across an entire utterance. The \textbf{\textit{s-vectors}} model is also trained on the same data (Voxceleb) as the \textbf{\textit{x-vectors}} model.

\vspace{-1.0em}

\subsection{Integrating speaker vectors into the system}

Speaker information is known to us while training the model. Hence during training, we make use of speaker-wise \textbf{\textit{x-vectors}} and \textbf{\textit{s-vectors}}. On the other hand, while testing, since speaker information is not known a priori, we apply utterance-wise speaker vectors. We now describe two ways of incorporating the speaker vectors.

\vspace{-0.5em}
\subsubsection{Concatenate speaker vectors to the encoder input}
\label{sec:Concatenate}
\begin{figure}[ht!]
    \centering
    \scalebox{0.8}{
	\includegraphics[width=1.0\columnwidth]{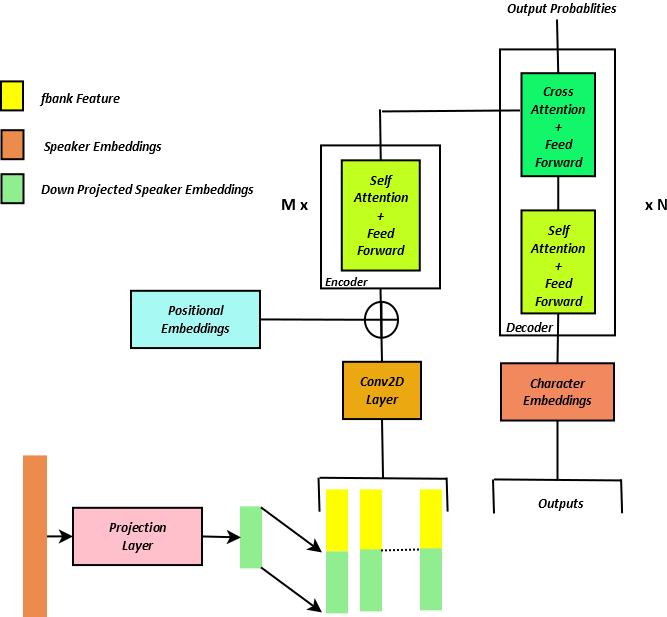}}
	\caption{Concatenating down projected speaker embeddings to acoustic feature frames}
\label{fig:Concat}
\end{figure}

Given an utterance, all the feature frames belong to one speaker. Hence each frame from a given utterance is concatenated with the \textbf{\textit{same}} down projected speaker vector. Here the 512 dimension speaker vector is first down projected to 83 and then concatenated to the 83-dimensional input acoustic frames as shown in Figure \ref{fig:Concat}. The input to the conv2d has the dimension 83 + 83 = 166. The down projection is learned during training. This model is referred to as \textit{x-vector\_cat} and \textit{s-vector\_cat} in this paper. Later in section \ref{sec:Specaug_norm} we describe the benefits of SpecAug \cite{specaugment} on these speaker vectors.

\vspace{-1.5em}

\subsubsection{Add speaker vectors to the encoder input}

In this approach, to every frame from the given utterance, the corresponding speaker vector is added. As discussed in section \ref{sec:Concatenate}, the 512 dimension speaker vector is first down projected to 83 to match the input fbank feature vectors' dimension, and then both of them are added.  Like in the concatenate case, the down projection is learned during training. This model is referred to as \textit{x-vector\_add}, and \textit{s-vector\_add} in this paper. Note the added vectors are first passed through two conv2D blocks before passing to encoder blocks.

\section{Data set Details}
\subsection{NPTEL Data Set}
National Programme on Technology Enhanced Learning (NPTEL\footnote{https://nptel.ac.in/}) is an open-source e-learning portal managed and coordinated by the Indian Institutes of Technology (IITs) and Indian Institute of Science (IISc). NPTEL provides free access to lecture videos and study materials for courses taught at top Indian universities. 

The lecture video and its corresponding subtitles have been made available in the public domain. In this paper, we have worked with lectures from Humanities, Computer Science, Electrical Engineering, and Mechanical Engineering domains. The audio was extracted from .mp4 lecture video files. The transcriptions were obtained from their corresponding subtitles (SRT files). The speaker, i.e., the instructor for each course, is different. All lectures are in Indian English. 
Note that, an NPTEL course runs over an entire semester. 
Given that the conditions, both external (like classroom conditions) and internal (like the instructors' health, pace etc.), vary, the NPTEL speech data has enough variability for a given speaker. The NPTEL dataset details are given in Table \ref{tab:NPTEL_dataset_stat}. Here the NPTEL-tr, NPTEL-dt and, NPTEL-et comprise data from 273 speakers. The train set NPTEL-tr has about 1 - 1.5 hours of data per speaker, while NPTEL-dt and NPTEL-et have these speakers contributing not more than five minutes of speech. There is no speech data overlap among ``tr,'' ``dt,'' and et''. Additional evaluation sets NPTEL-Unseen\_1 and NPTEL-Unseen\_2 are taken, and they comprise data from 50 unseen speakers, i.e., different from the 273 speakers present in NPTEL-tr, NPTEL-dt and, NPTEL-et. The unseen data was taken from 50 NPTEL courses that were not used while creating ''tr,'', ``dt,'' and ``et'' sets. These sets also have not more than five minutes of speech for each of the 50 unseen speakers. It is the same 50 speakers present in both NPTEL-Unseen\_1 and NPTEL-Unseen\_2.

\begin{table}[ht!]
\centering
\scalebox{0.8}{
\begin{adjustbox}{width=\columnwidth,center}
\begin{tabular}{lcc}
\hline
                         & \multicolumn{1}{l}{\textbf{Duartion(hrs)}} & \multicolumn{1}{l}{\textbf{\#Speakers}} \\ \hline \hline
\textbf{NPTEL-tr}        & 358.0                                       & 273                                      \\ 
\textbf{NPTEL-dt}        & 9.9                                         & 273                                      \\ 
\textbf{NPTEL-et}        & 9.8                                         & 273                                      \\ 
\textbf{NPTEL-Unseen\_1} & 2.0                                         & 50                                       \\ 
\textbf{NPTEL-Unseen\_2} & 2.0                                         & 50                                       \\ \hline
\end{tabular}
\end{adjustbox}}
\caption{NPTEL data set statistics}
\label{tab:NPTEL_dataset_stat}
\end{table}

\vspace{-2.5em}

\subsection{Librispeech Data set}
From the Librispeech Data set \cite{librispeech_Dataset}, we use the ``train\_other\_500'' split to train our models. This train set has data from 1166 speakers, totaling 496.7 hours of speech data. Each speaker in the train set has contributed about 30 minutes of speech data. We report our results on ``dev\_clean,'' ``dev\_other,'' ``test\_clean'' and, ``test\_other'' test sets. Here the suffix ``other'' is used to refer to a relatively noisy data set. 

\section{Experimental Details}
\label{sec:parameters}
The sequence to sequence encoder-decoder model is trained using the transformer \cite{attention_is_all_you_need} framework. Our NPTEL model encoder has 12 layers, whereas the decoder has six layers. Each layer in both the encoder and decoder has a self-attention module followed by 2048 feed-forward units with Rectified Linear Units (ReLU) as non-linearity. The self-attention module has an attention dimension of 256 and four attention heads. The batch size has been set to 32 with accumgrad 1.Noam optimizer (section 5.3 of \cite{attention_is_all_you_need}) with 25000 warmup steps and an initial transformer learning rate of 5 is used. All the models are trained for 40 epochs. The final model is obtained by averaging the best three models. In the Librispeech experiments, instead of batch-size, batch-bin was set to 8M and accumgrad was set to 8. All the remaining parameters were the same as the NPTEL experiments.
All the experiments were run on the 2080Ti and A100 GPU cards. Kaldi \cite{kaldi_toolkit} and ESPnet \cite{espnet_toolkit} tool kits were used to train our models. The model training was based on the hybrid CTC/Attention \cite{hybridCTC_ATTENTION} architecture with CTC contribution set to 0.3 while training and decoding. In this experiment, we use Byte Pair Encoding (BPE) \cite{BPE} subword units.

\section{Results and Analysis}

\subsection{NPTEL Data set results}
\label{sec:NPTEL_results}
Given in Table \ref{tab:NPTEL} are the percentage Word Error Rates (WERs) on the four NPTEL test sets. It can be seen that while concatenation helps, adding speaker embedding degrades the performance. On average, \textit{s\_cat} gives an absolute improvement of 1.5\% over the baseline on both seen speakers and unseen speaker scenarios. We see that \textbf{\textit{s-vectors}} are consistently better than the \textbf{\textit{x-vectors}} on adding and concatenating to the input fbank feature vectors. 

\begin{table}[ht!]
\centering
\scalebox{1.0}{
\begin{adjustbox}{width=\columnwidth,center}
\begin{tabular}{cccccc}
\hline
\multicolumn{1}{l}{} & \textbf{Baseline} & \textbf{x\_add} & \textbf{s\_add} & \textbf{x\_cat} & \textbf{s\_cat} \\ \hline \hline
\textbf{NPTEL-dt}           & 22.09             & 22.82             & 22.50              & 20.62             & 20.42             \\ 
\textbf{NPTEL-et}          & 21.93             & 22.89             & 22.22             & 20.50              & 20.26             \\ 
\textbf{NPTEL-Unseen\_1}     & 23.94             & 24.59             & 24.51             & 23.15             & 22.27            \\ 
\textbf{NPTEL-Unseen\_2}     & 24.37             & 25.21             & 25.14             & 23.19             & 22.62             \\ \hline
\end{tabular}
\end{adjustbox}}
\caption{NPTEL data set: Average error rate (\%WER)}
\label{tab:NPTEL}
\end{table}
\vspace{-1.5em}
We performed an analysis of speaker adaptation performance based on the utterance duration. In the NPTEL train set,  25\% of the utterances were of duration lesser than 5 seconds. The utterance duration-wise distribution of NPTEL train data is given in Table \ref{tab:NPTEL_Split}.

\begin{table}[ht!]
\centering
\scalebox{0.70}{
\begin{adjustbox}{width=\columnwidth,center}
\begin{tabular}{cccc}
\hline
\multicolumn{1}{l}{} & \multicolumn{1}{l}{\textbf{less\_5}} & \multicolumn{1}{l}{\textbf{5\_above}} \\ \hline \hline 
\textbf{NPTEL-tr}         & 23.0\%                & 77.0\%              \\
\textbf{NPTEL-dt}         & 25.0\%                & 75.0\%              \\
\textbf{NPTEL-et}         & 24.0\%                & 76.0\%              \\
\textbf{NPTEL-Unseen\_1}  & 21.0\%                & 79.0\%              \\
\textbf{NPTEL-Unseen\_2}  & 23.0\%                & 77.0\%              \\\hline 
\end{tabular}
\end{adjustbox}}
\caption{Percentage wise data split based on utterance duration in NPTEL data set}
\label{tab:NPTEL_Split}
\end{table}
\vspace{-1.5em}
Like in the NPTEL train set, the test sets also had most utterances with a duration of more than 5 seconds as seen in Table \ref{tab:NPTEL_Split}.

To analyze better, we split the test sets into two parts based on the duration discussed in Table \ref{tab:NPTEL_Split} and re-scored them. The average \%WERs on the two splits are given in Tables \ref{tab:NPTEL_five} and \ref{tab:NPTEL_five_above}. 
Here the suffixes ``less\_5'' and ``5\_above'' 
are used to specify the test set utterance duration as discussed in Table \ref{tab:NPTEL_Split}.

\begin{table}[ht!]
\centering
\scalebox{1.0}{
\begin{adjustbox}{width=\columnwidth,center}
\begin{tabular}{cccccc}
\hline
\textbf{}                & \textbf{Baseline} & \textbf{x\_add} & \textbf{s\_add} & \textbf{x\_cat} & \textbf{s\_cat} \\ \hline \hline
\textbf{NPTEL-dt - less\_5}      & 27.1              & 28.0                & 27.9              & 25.8              & 25.6              \\ 
\textbf{NPTEL-et - less\_5}     & 26.1              & 27.2              & 26.9              & 25.1              & 24.7              \\ 
\textbf{NPTEL-Unseen\_1 - less\_5} & 26.3              & 27.2              & 26.6              & 25.3              & 24.9              \\ 
\textbf{NPTEL-Unseen\_2 - less\_5} & 27.0              & 28.6              & 26.9              & 25.6              & 25.7              \\ \hline
\end{tabular}
\end{adjustbox}}
\caption{NPTEL data set: Average error rate (\%WER) on utterances with duration less than five seconds}
\label{tab:NPTEL_five}
\end{table}
\vspace{-1.0em}

\begin{table}[ht!]
\scalebox{1.0}{
\begin{adjustbox}{width=\columnwidth,center}
\begin{tabular}{cccccc}
\hline
\textbf{ }           & \textbf{Baseline} & \textbf{x\_add} & \textbf{s\_add} & \textbf{x\_cat} & \textbf{s\_cat} \\ \hline \hline
\textbf{NPTEL-dt - 5\_above}        & 21.0              & 21.7              & 21.3  & 19.5              & 19.3                            \\ 
\textbf{NPTEL-et - 5\_above}        & 21.1              & 22.0              & 21.3 & 19.6              & 19.4                            \\ 
\textbf{NPTEL-Unseen\_1 - 5\_above} & 23.5              & 24.1              & 24.1 & 22.8              & 21.8                            \\ 
\textbf{NPTEL-Unseen\_2 - 5\_above} & 23.8              & 24.5              & 24.8 & 22.7              & 22.0                            \\ \hline
\end{tabular}
\end{adjustbox}}
\caption{NPTEL data set: Average error rate (\%WER) on utterances with duration greater than 5 seconds}
\label{tab:NPTEL_five_above}
\end{table}
\vspace{-1.0em}

\vspace{-1.0em}
From Tables \ref{tab:NPTEL_five} and \ref{tab:NPTEL_five_above}, we observe that both \textbf{\textit{x-vector}} and \textbf{\textit{s-vector}}, when concatenated, improve over the baseline, validating the benefit of adaptation techniques across test splits. It is worth noting that the \textit{x\_cat} and \textit{s\_cat} perform better than the baseline on ``NPTEL-Unseen\_1'' and ``NPTEL-Unseen\-2'', i.e., unseen test sets as well. In the NPTEL database, the train and test sets distribution based on utterance duration was similar. This, we believe, leads to a consistent pattern in the results across both splits (``less\_5'' and ``5\_above''). 

\subsection{Librispeech Dataset results}
\label{sec:Libri_results}
Next, we investigate the speaker adaptation performance on Librispeech. We repeat the experiments discussed in the previous section on the 500-hour Librispeech train ``other'' set. This is a comparatively noisy set, and unlike the NPTEL data set where we had about 1 - 1.5 hours of speech data per speaker, the 500-hour train set of Librispeech has only 30 minutes of speech data per speaker.

\begin{table}[ht!]
\centering
\scalebox{0.9}{
\begin{adjustbox}{width=\columnwidth,center}
\begin{tabular}{cccccc}
\hline
\multicolumn{1}{l}{} & \multicolumn{1}{l}{\textbf{Baseline}} & \textbf{x\_add} & \textbf{s\_add} & \textbf{x\_cat} & \textbf{s\_cat} \\ \hline \hline
\textbf{dev\_clean}    & 6.8                                    & 6.3               & 6.4               & 6.2               & 6.7               \\ 
\textbf{test\_clean}   & 6.8                                    & 6.1               & 6.2               & 6.3               & 6.5               \\ \hline 
\textbf{dev\_other}    & 13.8                                   & 14.8              & 14.1              & 15.1              & 15.5              \\ 
\textbf{test\_other}   & 14                                     & 14.5              & 14.3              & 15.3              & 15.6              \\ \hline
\end{tabular}
\end{adjustbox}}
\caption{Librispeech: Average error rate (\%WER)}
\label{tab:Librispeech}
\end{table}
In Table \ref{tab:Librispeech}, the \%WERs on the Librispeech test sets are given. We observe that while we get an improvement in performance across all speaker adapt approaches on the ``clean'' test sets, there is a degradation in the case of the ``other'' test sets. The other set is relatively noisy, and \textbf{\textit{x-vectors/s-vectors}} are estimated from them. However, \textbf{\textit{x-vectors/s-vectors}}  extractors have been trained on relatively clean voxceleb data. We investigated this behavior by performing an in-depth analysis based on utterance duration akin to what we did in the NPTEL case to investigate the impact of the discussed adaptation techniques in a scenario wherein the utterance duration-wise distribution is not the same between train and test sets.

\begin{table}[ht!]
\centering
\scalebox{0.8}{
\begin{adjustbox}{width=\columnwidth,center}
\begin{tabular}{cccc}
\hline
\multicolumn{1}{l}{} & \multicolumn{1}{l}{\textbf{less\_5}} & \multicolumn{1}{l}{\textbf{5\_15}} & \multicolumn{1}{l}{\textbf{15\_above}} \\ \hline \hline
\textbf{train\_other\_500}   & 9\%          & 65\%        & 26\%        \\ 
\textbf{dev\_clean} & 41.0\%    &52.0\%  &7.0\%   \\
\textbf{test\_clean} & 42.0\% & 49.0\% & 9.0\% \\ \hline 
\textbf{dev\_other} & 47.0\% &48.0\% & 5.0\% \\
 \textbf{test\_other} & 48.0\% & 46.0\% & 6.0\% \\ \hline
\end{tabular}
\end{adjustbox}}
\caption{Percentage wise data split based on utterance duration in Librispeech train set}
\label{tab:Librispeech_Split}
\end{table}

\vspace{-1.5em}
The majority (i.e., 91\%) of the Librispeech train sets' utterances had a duration greater than 5 seconds. The utterance duration wise distribution of train data is given in Table \ref{tab:Librispeech_Split}. As stated before, unlike in the case of the NPTEL data set, the Librispeech test sets followed a different distribution of utterance duration. Even though majority of utterances in the train set had duration greater than 5 seconds, that was not the case in the test sets. The utterance duration-wise data split of the Librispeech test sets is also given in Table \ref{tab:Librispeech_Split}.

The results on the three test split discussed in Table \ref{tab:Librispeech_Split} are given in Tables \ref{tab:Librispeech_five}, \ref{tab:Librispeech_five_fifteen} and, \ref{tab:Librispeech_fifteen}. The effect of lack of utterances with a duration lesser than five seconds in the train set reflects the performance of proposed adaptation approaches on test split ``less\_5''. As observed in Table \ref{tab:Librispeech_five}, all the adaptation techniques perform on par (or slightly worse) with the baseline in the case of  ``clean'' and are always worse on the ``other'' sets. The ``other'' sets being comparatively noisy are worst affected. The speaker embedding estimated from very short noisy utterances seems to fail to capture the speaker characteristics well.

\begin{table}[ht!]
\scalebox{1.0}{
\begin{adjustbox}{width=\columnwidth,center}
\begin{tabular}{cccccc}
\hline
\multicolumn{1}{l}{}      & \multicolumn{1}{l}{\textbf{Baseline}} & \textbf{x\_add} & \textbf{s\_add} & \textbf{x\_cat} & \textbf{s\_cat} \\ \hline \hline
\textbf{dev\_clean - less\_5}   & 7.2                                    & 7.2               & 7.5               & 7.2               & 7.7               \\ 
\textbf{test\_clean - less\_5} & 7.4                                    & 7.9               & 7.8               & 7.8               & 8.3               \\ \hline 
\textbf{dev\_other - less\_5}  & 14.2                                   & 16.2              & 15.5              & 16.1              & 16.4              \\ 
\textbf{test\_other - less\_5} & 15.5                                   & 17.9              & 17.5              & 18.3              & 18.7              \\ \hline
\end{tabular}
\end{adjustbox}}
\caption{Librispeech: Average error rate (\%WER) on utterances with duration lesser than five seconds }
\label{tab:Librispeech_five}
\end{table}
\vspace{-1.5em}
As we move to the ``5\_15'' split, we notice that the adaptation techniques start to catch up with the baseline even in the ``other'' scenario. The results on the ``5\_15'' split are given in the Table \ref{tab:Librispeech_five_fifteen}. The performance improvement can be attributed to two factors. 1) Considerable amount of train utterances in this duration range, and 2) As the utterances get longer, the speaker vectors, both \textbf{\textit{x-vectors}}, and \textbf{\textit{s-vectors}} are calculated better because of better averaging during stats pooling.

\begin{table}[ht!]
\scalebox{1.0}{
\begin{adjustbox}{width=\columnwidth,center}
\begin{tabular}{cccccc}
\hline
\multicolumn{1}{l}{}      & \multicolumn{1}{l}{\textbf{Baseline}} & \textbf{x\_add} & \textbf{s\_add} & \textbf{x\_cat} & \textbf{s\_cat} \\ \hline \hline
\textbf{dev\_clean - 5\_15}  & 5.9                                    & 5.8               & 5.9               & 5.8               & 6.2               \\ 
\textbf{test\_clean - 5\_15} & 5.8                                    & 5.3               & 5.5               & 5.7               & 5.9               \\ \hline 
\textbf{dev\_other - 5\_15}  & 13.3                                   & 14.2              & 13.5              & 14.7              & 15.2              \\ 
\textbf{test\_other - 5\_15} & 13.1                                   & 13.6              & 13.3              & 14.6              & 15                \\ \hline
\end{tabular}
\end{adjustbox}}
\caption{Librispeech: Average error rate (\%WER) on utterances with duration between 5 and 15 seconds }
\label{tab:Librispeech_five_fifteen}
\end{table}
\vspace{-1.0em}
Presence of a sufficient number of utterances with a duration of over 15 seconds while training (Table \ref{tab:Librispeech_Split}) and the benefit obtained from having longer utterances can be better observed in Table \ref{tab:Librispeech_fifteen}. We can see that while being better on the ``clean'' set, all the adaptation methods perform better than the \textit{Baseline} on the ``other'' set as well. This behavior, as discussed in the previous paragraph, we believe is because of better-averaged \textbf{\textit{x-vector}} and \textbf{\textit{s-vector}} obtained from the long noisy test utterances.
\vspace{-1.5em}
\begin{table}[ht!]
\scalebox{1.0}{
\begin{adjustbox}{width=\columnwidth,center}
\begin{tabular}{cccccc}
\hline
\multicolumn{1}{l}{}     & \multicolumn{1}{l}{\textbf{Baseline}} & \textbf{x\_add} & \textbf{s\_add} & \textbf{x\_cat} & \textbf{s\_cat} \\ \hline \hline
\textbf{dev\_clean - 15\_above}  & 9                                      & 7.2               & 6.4               & 6.4               & 6.8               \\ 
\textbf{test\_clean - 15\_above} & 8.4                                    & 6.4               & 6.2               & 6.4               & 6.1               \\ \hline 
\textbf{dev\_other - 15\_above}  & 15                                     & 14.2              & 13.5              & 14.7              & 15                \\ 
\textbf{test\_other - 15\_above} & 14.3                                   & 12.2              & 12.6              & 12.8              & 13.2              \\ \hline
\end{tabular}
\end{adjustbox}}
\caption{Librispeech: Average error rate (\%WER) on utterances with duration greater than 15 seconds }
\label{tab:Librispeech_fifteen}
\end{table}

\vspace{-2.5em}

\subsection{Effects of SpecAugment and Normalization of speaker embedding}
\label{sec:Specaug_norm}
SpecaugAugment \cite{specaugment} has been shown to provide effective regularization to ASR training. Given a spectrogram, SpecAugment is used to make the ASR system robust to distortions in frequency direction and loss of speech segments in time direction. In all the experiments discussed so far, SpecAugment was used both on fbank features as well as speaker vectors. 
In the data sets we have used, an utterance belongs to a single speaker. Hence the same constant speaker embedding will be added or concatenated to all the feature frames of that utterance. Hence, to bring about a regularizing effect, we SpecAugment the speaker embeddings along with fbank features. At the very input to the model, the 83 dimensional Cepstral Mean and Variance Normalized (CMVN) fbank features are appended with 512 dimension speaker embeddings. That is, at the input of the encoder, we have an 83 (CMVN fbank dim) + 512 (speaker embedding dim) = 595 dimension vector for every frame in a given utterance. Since SpecAugment is the first step to be executed, it considers the 595 dimension vector at the input while time/frequency masking and time warping. Post SpecAugmnet, based on the adaptation approach we wish to follow, we separate the 512 dimension speaker embedding, down project it to 83, and then add it or concatenate it to the fbank vectors.  

Another important factor to be taken care of is that the speaker embedding values and the CMVN fbank feature vector values have a different dynamic range. So we need to ensure that we length normalize the 512 dimension speaker embeddings. While training the model, we deal with a batch of examples. If in a batch there are ``B'' examples, each having ``T'' frames, then at the input, we have ``BxTx595'' dimension matrix, in which the lower half, i.e., ``BxTx512'' is the speaker embedding matrix. This matrix can be length normalized along the batch axis, i.e., dimension ``B'', or along the frames axis, i.e.,``T'' or along the frequency bin access, i.e., along ``512'' dimensions. We call them ``B\_Norm,'' ``T\_Norm,'' and ``F\_norm'' respectively. Here we have employed length (i.e., $L_{2} $ norm) normalization.

In Table \ref{tab:Specaug_Norm_Analysis}, given are the Average \%WERs on ``dev\_clean'' and ``test\_clean'' sets from Librsipeech. Here ``WS'' and ``NS'' refer to ``With SpecAugmentation,'' and ``No SpecAugmentation'' respectively. ``B\_Norm,'' ``T\_Norm,'' and ``F\_norm'' are as discussed in the preceding paragraph. To train the model for this experiment, we have used the ''train\_clean\_360'' subset of Librsipeech. Since we are training using the clean set, the results are reported only on the clean test sets.``Baseline'' refers to the baseline experiment without Speaker embeddings. In the ``Baseline'' experiment, ``WS'' and ``NS'' refer to application of SpecAugment on the fbank features.

\begin{table}[ht!]
\centering
\scalebox{0.8}{
\begin{adjustbox}{width=\columnwidth,center}
\begin{tabular}{lcc}
\hline
                                 & \textbf{dev\_clean} & \textbf{test\_clean} \\ \hline \hline
\textbf{Baseline - With SpecAug(WS)}             & 6.8                 & 7                    \\ 
\textbf{Baseline - No SpecAug (NS)}             & 7.1                 & 7.4                  \\ 
\textbf{x\_cat - NS - B\_Norm} & 7.5                 & 7.8                  \\ 
\textbf{x\_cat - NS - F\_Norm} & 7.2                 & 7.7                  \\ 
\textbf{x\_cat - NS - T\_norm} & 7.4                 & 7.4                  \\ 
\textbf{x\_cat - WS - T norm}  & 6                   & 6.3                  \\ \hline
\end{tabular}
\end{adjustbox}}
\caption{Average error rate (\%WER) by applying SpecAugment and Normalization on speaker embeddings}
\label{tab:Specaug_Norm_Analysis}
\end{table}

\vspace{-1.5em}

The benefit of SpecAugment is evident from the \textit{Baseline-WS} and \textit{Baseline-NS} results. We considered the \textit{x-in cat} experiment as a reference and analyzed the effects of different normalization approaches. We can see that both ``F\_Norm'' and ``T\_Norm'' results are competitive to each other. This was observed in our other ``cat'' experiments as well. We chose ``T\_Norm'' and performed an experiment along with SpecAugment and obtained a significant reduction in \%WER over the baseline with SpecAugment. Hence, in all our adaptation experiments discussed in section \ref{sec:NPTEL_results}, and section \ref{sec:Libri_results} we have used ''T\_Norm'' with SpecAugment.
\vspace{-1.0em}
\section{Conclusion}
This paper has investigated two different approaches to improve transformer-based ASR models’ recognition performance by incorporating speaker information while training. Two types of speaker embeddings: standard \textbf{\textit{x-vectors}} and \textbf{\textit{svectors}} proposed in our previous work, were used. We analyze the effect of utterance duration on the discussed adaptation techniques. SpecAugmentation and different normalization approaches in the speaker embeddings were also discussed. Our experiments show that use of speaker embeddings helps improve the speaker adaptation performance by a significant margin.

\bibliographystyle{IEEEtran}
\bibliography{template}

\end{document}